\documentclass[iop]{emulateapj}
\usepackage[T1]{fontenc}
\usepackage{ae,aecompl}
\usepackage{graphicx}   % Including figure files
\usepackage{amsmath}    % Advanced maths commands
\usepackage{amssymb}    % Extra maths symbols
\usepackage{physics}
\usepackage{url}
\usepackage{threeparttable}
\usepackage{upgreek}
\newcommand{\mdeg}{\ensuremath{^{\circ}}}
\newcommand{\ppcc}{$\,$pc$\,$cm$^{-3}$} % parsec per cm-cubed
\shorttitle{}
\shortauthors{Yogesh Maan et al.}
\begin{document}

\title{A Search for Pulsars in Steep Spectrum Radio Sources}
%\correspondingauthor{Yogesh Maan}
\author{Yogesh Maan}
\email{maan@astron.nl}
\author{Cees Bassa}
\author{Joeri van Leeuwen\altaffilmark{1}}
\affil{ASTRON, the Netherlands Institute for Radio Astronomy, Postbus 2, 7990 AA, Dwingeloo, The Netherlands}

\and

\author{M. A. Krishnakumar\altaffilmark{2}}
\author{Bhal Chandra Joshi}
\affil{National Centre for Radio Astrophysics, Tata Institute of Fundamental Research, Post Bag 3, Ganeshkhind, Pune - 411007, India}
\altaffiltext{1}{Astronomical Institute ``Anton Pannekoek'', University of Amsterdam, Science Park 904, 1098 XH Amsterdam, The Netherlands}
\altaffiltext{2}{Radio Astronomy Centre, NCRA-TIFR, Udagamandalam, India}
\begin{abstract}
We report on a time-domain search for pulsars in 44 steep spectrum radio sources originally identified from recent imaging surveys. The time-domain search was conducted at 327\,MHz using the Ooty radio telescope, and utilized a semi-coherent dedispersion scheme, retaining the sensitivity even for sub-millisecond periods up to reasonably high dispersion measures. No new pulsars were found. We discuss the nature of these steep spectrum sources and argue that majority of the sources in our sample should either be pulsars or a new category of Galactic sources. Several possibilities that could hinder detection of these sources as pulsars, including anomalously high scattering or alignment of the rotation and magnetic axes, are discussed in detail, and we suggest unconventional search methods to further probe these possibilities.
\end{abstract}
%%%%%%%%%%%%%%%%%%%%%%%%%%%%%%%%%%%%%%%%%%%%%%%%%%
%%%%%%%%%%%%%%%%%%%%%%%%%%%%%%%%%%%%%%%%%%%%%%%%%%
\section{Introduction}
Pulsars are one of the few classes of astronomical sources that have steep
radio spectra. In fact, many pulsars have been identified after noticing
the steep spectra of their counterparts in continuum images.
In many cases, this led to break through discoveries or unearthed especially interesting systems.
It was through the unusual, steep spectral properties of the continuum source 4C~21.53W that
the first millisecond pulsar (MSP) PSR~B1937+21 was discovered \citep{Backer82}.
Similarly, the discovery of the first ever pulsar
in a globular cluster \citep[{PSR~B1821$-$24} in M28;][]{HHB85,Lyne87} was
strongly motivated by its steep spectrum.
{PSR~J0218+4232} was also first detected as a steep spectrum source in
continuum imaging using the Westerbork synthesis radio telescope before
discovery of the associated radio pulsations \citep{Navarro95}. It
turned out to be the farthest MSP known in the field at that time. 
Some other notable cases where prior information about the steep spectrum
assisted the pulsar discovery are: PSR~J2022+3842,
the central source of supernova remnant G76.9+1.0,
concurrently detected in radio imaging and time-domain by
\citet[][]{Marthi11}, independently from the X-ray discovery by \citet{2011ApJ...739...39A};
and PSR~J0815+4611, which was first identified as a point source in the deep
epoch-of-reionization observations of the 3C196 field and turned out to be
a highly polarized nearby pulsar (Kondratiev et al., private communication)\footnote{See
\url{http://www.astron.nl/dailyimage/main.php?date=20150331}
and \url{http://www.astron.nl/lotaas/}}.
More recently, image-based searches utilizing steep-spectrum as one of the
main criteria have uncovered 8 new pulsars from \textsl{Fermi} Large Area
Telescope unidentified sources \citep{Bhakta17,Frail18}. These
discoveries demonstrate the potential of spectral information based searches
for radio pulsars, and motivate investigation of the steep spectrum sources
for the presence of any pulsations.
\par
There have not been many surveys of steep spectrum sources in past, partly
because a few that were undertaken were not very successful in discovering
new pulsars. For example, \citet{Damico85} searched 18 steep radio sources
for presence of pulsars, however, their sampling time of 0.25\,ms would have
reduced the sensitivity particularly for fast pulsars. \citet{Kaplan00}
found 16 compact (typical angular size $<0^{\prime\prime}.2$), steep spectrum
sources, but suggested that most of those are perhaps high-redshift galaxies.
\citet{Crawford00} did a systematic search for pulsars in 92 significantly
polarized, unidentified compact sources but did not find any pulsations. While
these surveys were sensitive enough to detect the associated pulsars with high
significance, the available observing
capabilities \citep[e.g., coarse sampling time as mentioned above, or
scintillation-prone narrow bandwidth of 1\,MHz in][]{Crawford00}
at those times could have affected the results for some of the targets.
Moreover, techniques to detect exotic sources more effectively (e.g.,
acceleration searches) and the compute resources to use such techniques
at large scales have become available only more recently, and conducting
such pulsars surveys now could be more fruitful. 
\par
Sensitive all sky imaging surveys have been uncovering interesting new
sources, including ones with steep spectra.
A recent 147\,MHz survey using the Giant Metrewave Radio Telescope (GMRT),
called the TIFR GMRT Sky Survey (TGSS), covers the whole sky north of
declination ($\delta$) $-53\mdeg$ \citep[90 percent of total sky;][]{Sirothia14}.
Some of the early results from this survey were presented in
\citet{Bagchi11,GopalKrishna12,Sirothia14,Krishna14}. The first
alternative data release of this survey \citep[TGSS~ADR1;][]{Intema17}
catalogs 0.62\,million radio sources above $7\sigma$ level.
TGSS~ADR1
in combination with earlier high radio frequency all sky surveys
\citep[e.g., the NRAO VLA Sky Survey, hereafter NVSS, at 1.4\,GHz;][]{Condon98},
presents a wonderful opportunity to study the spectral index distribution
of the astronomical sources in general, and find compact, steep
spectrum sources that are potential pulsar candidates in particular.
\par
\citet{deGasperin18} used TGSS~ADR1 and NVSS to make a spectral
index catalogue for nearly 80~percent of the whole sky and confirmed
an intriguing excess of compact and steep spectrum (with the spectral
index\footnote{Spectral index $\alpha$ is defined via the conventional
power law representation of the spectrum: $S\propto\nu^\alpha$, where
$S$ and $\nu$ are the flux density and observing frequency, respectively.}
$\alpha$$<$$-1.5$) sources in the Galactic plane
first indicated by \citet{DeBreuck00}. Since pulsars are well known for
their steep spectra and there is no other known
category of compact and such steep spectrum sources in the Galaxy, the
above excess makes a compelling case to search for radio pulsations
from the Galactic steep spectrum sources. While some of these sources
are bright enough that pulsations should have been detected in earlier
high frequency pulsar surveys, many are faint and some are detected only
in TGSS. In any case, deep and well designed pulsation searches are needed
to detect or rule out exotic classes of pulsars (e.g., highly relativistic binaries,
sub-millisecond pulsars, etc.) possibly harboring these sources, before
considering a new class of Galactic steep spectrum sources. With these
motivations, we have selected two groups of steep spectrum sources
as potential pulsar candidates and searched for any pulsations from
them. To minimize the effects of interstellar scattering as well as
use the advantage of intrinsically larger flux density at lower
frequencies, we have used a frequency of 327\,MHz as an optimal
choice for our survey.
\par
Details of the sample selection, observations, and our data
reduction and search methods are presented in the next section.
Section~3 presents the results of our search, Section~4 comprises
a detailed discussion on the nature of sources in our sample,
followed by a summary in the last Section.
\section{Target sources and methods}
\subsection{Sample Selection}
\citet{Tiwari16} and \citet{deGasperin18} have combined data at 147\,MHz
and 1.4\,GHz from TGSS and NVSS, respectively, to make spectral index
maps as well as source catalogues for nearly 80~percent of the whole sky
(assuming no variability between the epochs of the two surveys).
These catalogues have provided useful starting points for our selection
of compact, steep spectrum sources. Specifically, we have chosen two
samples: sources within Galactic plane and targets with steepest known
spectra. More details of these samples are given below.
\subsubsection{The Galactic-plane (GP) sample}
We first chose all the sources from \citet{Tiwari16} that are
detected with a significance of more than 10\,$\sigma$ ($\gtrsim$35\,mJy)
in TGSS,
and have spectral
indices ($\alpha$) steeper than $-1.4$, declination in the range $-45$\mdeg\ to
$+45$\mdeg\ (a constraint from observing setup, see next
subsection), and angular sizes less than 50$^{\prime\prime}$. \citet{Intema17}
also define the observed multiplicity of the source structure in terms of
Gaussian components. To increase the chance of retaining only point-like sources,
we also discarded the ones with complex structures (i.e., the ones which need
multiple or overlapping Gaussians to fit) or whose fitted positions differ by
more than 20$^{\prime\prime}$ in TGSS and NVSS. The sample was then filtered out
for any already known identifications using NASA/IPAC extragalactic database
(NED). The remaining sources were (again) cross-matched with NVSS and their
spectral indices were verified. Towards the end of data reduction
for this sample, \citet{deGasperin18} published their spectral index catalog.
We also cross-matched with their catalog and found
our spectral indices to be consistent.
\par
Apart from pulsars, high redshift radio galaxies (HzRGs) constitute
another class of astronomical sources that exhibit steep radio spectra
and compact angular sizes \citep{MdB08}. Some HzRGs and quasars also tend to be bright
in infrared. 
In our ``Galactic-plane sample'', we try to
minimize the probability that the chosen source is a HzRG by confining
the selection to a narrow Galactic latitude (\emph{b}) range of $-2$~to~$+2$ and
looking at the infrared properties of the sources. We cross-matched the
sources with allWISE source catalog\footnote{\url{http://wise2.ipac.caltech.edu/docs/release/allwise/}}
and examined the allWISE image atlas
to exclude the targets with any obvious infrared counterparts. Then,
the remaining sources were cross-matched with the ATNF pulsar catalog
\citep{Manchester05} to identify any known pulsar.
A total of 3 known pulsars, including one in a Globular cluster, were
identified at various filtering stages mentioned above. Using the spectral
index catalog by \citet{deGasperin18}, we note that there are 16 more known
pulsars that are identifiable in TGSS~ADR1 within the Galactic lattude range
of $-2$ to $+2$ and which have spectral indices steeper than $-1.4$. However,
their high frequency counterparts are not detected in NVSS. Since our GP sample
has considered only the sources that have counterparts detected in both the
surveys, these 16 pulsars naturally did not get selected.
Finally, TGSS~ADR1
cut-out images\footnote{\url{http://tgssadr.strw.leidenuniv.nl}}
around each of the source were
examined to exclude any candidates due to imaging-artifacts.
The final Galactic-plane sample consists of 25 sources with spectral
indices in the range $-$1.80 to $-$1.44, and are listed
in Table~\ref{table_gp}.
\par

\subsubsection{The ultra-steep spectrum (USS) sample}
For this sample, we used catalogs from \citet{Tiwari16} and
\citet{deGasperin18} to select all sources having spectral indices steeper
than $-2.5$ and declination in the range $-45$\mdeg\ to $+45$\mdeg.
This sample consists of 21 sources, including 2 known pulsars
(see Table~\ref{table_uss}). We retained the 2 known pulsars in our
sample to use them as additional control pulsars, but do not consider
them in our discussion on the steep spectrum sources later.
Many of
these sources do not have an identifiable counterpart in NVSS implying
that the deduced spectral indices are in fact upper limits and actual
spectra could be even steeper. We note that the targets in this sample
were not scrutinized using many of the criteria detailed above for the GP~sample,
including examining the TGSS~ADR1 cut-out images to rule out any artifacts.
%%%%%%%
%%%=========================================================================
%%%=========================================================================
\begin{table*}
 \centering
  \caption{The Galactic plane sample: Observation details and upper limits on pulsed flux density.}
  \begin{threeparttable}
  \begin{tabular}{@{}rlcrrr@{}}
  \hline
  Sr. No. & Target$-$ID & $\alpha$ & $S_{\rm 325\,MHz}$ & $S_{\rm min}^{\rm pulsed}$ & $S_{\rm min}^{\rm aligned-pulsed}$ \\ 
          &             &          &  (mJy)             &   (mJy)            & (mJy) [\%] \\
  \hline

%%---------------------------------------%%
 1 & J050416+413334 &  $-$1.74$\pm$0.08 &    41.8 &          5.5 &    14.5 [ 34.7] \\
 2 & J051322+413839 &  $-$1.80$\pm$0.07 &    57.5 &          5.5 &    14.5 [ 25.2] \\
 3 & J054154+285045 &  $-$1.62$\pm$0.10 &    27.4 &          5.7 &    15.1 [ 55.4] \\
 4 & J054238+273955 &  $-$1.48$\pm$0.10 &    23.5 &          5.7 &    15.1 [ 64.6] \\
 5 & J054751+244742 &  $-$1.50$\pm$0.08 &    33.6 &          5.5 &    14.5 [ 43.2] \\
 6 & J055426+223641 &  $-$1.39$\pm$0.09 &    81.7 &          5.7 &    15.1 [ 18.5] \\
 7 & J055751+242623 &  $-$1.51$\pm$0.08 &    35.1 &          5.5 &    14.5 [ 41.3] \\
 8 & J060126+221705 &  $-$1.52$\pm$0.05 &   153.7 &          5.5 &    14.5 [ ~9.4] \\
 9 & J060633+240947 &  $-$1.49$\pm$0.06 &    60.2 &          5.5 &    14.5 [ 24.1] \\
10 & J061644+122121 &  $-$1.49$\pm$0.07 &    43.9 &          5.5 &    14.5 [ 33.0] \\
11 & J061726+181632 &  $-$1.72$\pm$0.07 &    56.5 &          5.5 &    14.5 [ 25.7] \\
12 & J061919+140306 &  $-$1.44$\pm$0.09 &    34.1 &          5.7 &    15.1 [ 44.4] \\
13 & J062043+114313 &  $-$1.62$\pm$0.08 &    37.3 &          5.5 &    14.5 [ 38.9] \\
14 & J062257+121518 &  $-$1.49$\pm$0.07 &    42.2 &          5.5 &    14.5 [ 34.4] \\
15 & J063148+102416 &  $-$1.70$\pm$0.09 &    34.4 &          5.2 &    13.9 [ 40.3] \\
16 & J064124+014858 &  $-$1.60$\pm$0.07 &    49.2 &          5.5 &    14.5 [ 29.5] \\
17 & J064801+032706 &  $-$1.52$\pm$0.07 &    44.8 &          5.5 &    14.5 [ 32.4] \\
18 & J185321+050802 &  $-$1.54$\pm$0.05 &    91.9 &         10.7 &    28.4 [ 30.9] \\
19 & J191120+073531 &  $-$1.51$\pm$0.06 &   105.0 &         13.1 &    34.7 [ 33.0] \\
20 & J191350+083553 &  $-$1.49$\pm$0.06 &    69.6 &         10.7 &    28.4 [ 40.8] \\
21 & J192014+111338 &  $-$1.55$\pm$0.06 &    89.9 &         10.7 &    28.4 [ 31.6] \\
22 & J200130+331242 &  $-$1.73$\pm$0.09 &    43.2 &          7.2 &    18.9 [ 43.9] \\
23 & J201759+363016 &  $-$1.52$\pm$0.10 &    39.3 &          9.5 &    25.2 [ 64.2] \\
24 & J202134+332733 &  $-$1.77$\pm$0.08 &    53.8 &          7.2 &    18.9 [ 35.2] \\
25 & J211746+470956 &  $-$1.70$\pm$0.07 &    49.8 &          6.4 &    17.0 [ 34.2] \\
%%---------------------------------------%%
\hline
\label{table_gp}
\end{tabular}
\begin{tablenotes}
\item[] Notes. --- (1) Target$-$ID denotes the source name based on the J2000
coordinates following the convention Jhhmmss+ddmmss.
(2) $S_{\rm 325\,MHz}$ is the expected continuum flux-density at 325\,MHz derived
using the TGSS flux density and the spectral index.
$S_{\rm min}^{\rm pulsed}$ is the formal upper-limit on pulsed flux-density
averaged over the period, assuming 10\% pulse duty-cycle.
(3) $S_{\rm min}^{\rm aligned-pulsed}$ is the upper-limit on pulsed flux-density
assuming a large pulse duty-cycle (e.g., due to scattering or for nearly
aligned rotators) of 70\% of the rotation period and
presented in mJy as well as a percentage of $S_{\rm 325\,MHz}$.
\end{tablenotes}
\end{threeparttable}
\end{table*}
%%%=========================================================================
\begin{table*}
 \centering
  \caption{The ultra-steep spectrum sample: Observation details and upper limits on pulsed flux density.}
  \begin{threeparttable}
  \begin{tabular}{@{}rlcrrr@{}}
  \hline
  Sr. No. & Target$-$ID & $\alpha$ & $S_{\rm 325\,MHz}$ & $S_{\rm min}^{\rm pulsed}$ & $S_{\rm min}^{\rm aligned-pulsed}$ \\
          &             &          &  (mJy)             &   (mJy)            & (mJy) [\%] \\
  \hline

%%---------------------------------------%%
 1 & J000559$-$010259                  &  $<-$2.59       &  $<$ 130 &          5.5 &    17.8 [ 13.7] \\
 2 & J011715$-$141154                  &  $-$2.67$\pm$0.11 &    190 &          5.9 &    19.3 [ 10.1] \\
 3 & J075034$-$164401                  &  $<-$2.50       &  $<$ 100 &          6.1 &    19.7 [ 19.7] \\
 4 & J080130+141712                    &  $<-$2.71       &  $<$ 140 &          5.8 &    18.8 [ 13.4] \\
 5 & J094135+153328                    &  $-$2.52$\pm$0.09 &    130 &          5.5 &    17.8 [ 13.7] \\
 6 & J122030+334532                    &  $<-$2.67       &  $<$ 110 &          5.5 &    17.8 [ 16.2] \\
 7 & J184723$-$040214 (PSR B1844$-$04) &  $<-$2.64       &  $<$ 120 &         16.8 &    54.3 [ 45.2] \\
 8 & J192145+215304  (PSR B1919+21)    &  $<-$2.88       &  $<$ 110 &          8.1 &    26.2 [ 23.8] \\
 9 & J194531+390704                    &  $<-$3.25       &  $<$ 490 &          6.7 &    21.7 [ ~4.4] \\
10 & J194954+375445                    &  $<-$3.22       &  $<$ 420 &          7.0 &    22.7 [ ~5.4] \\
11 & J194953+373843\tnote{$\dagger$}   &  $<-$2.69       &  $<$ 140 &          6.9 &    22.2 [ 15.9] \\
12 & J194956+374650                    &  $<-$2.92       &  $<$ 160 &          6.9 &    22.2 [ 13.9] \\
13 & J194956+374146\tnote{$\dagger$}   &  $<-$2.82       &  $<$ 160 &          6.9 &    22.2 [ 13.9] \\
14 & J194957+373236\tnote{$\ddagger$}  &  $<-$2.67       &  $<$ 110 &          7.0 &    22.7 [ 20.6] \\
15 & J194959+373423\tnote{$\ddagger$}  &  $<-$2.73       &  $<$ 150 &          7.0 &    22.7 [ 15.1] \\
16 & J195002+373822\tnote{$\dagger$}   &  $<-$2.97       &  $<$ 380 &          6.9 &    22.2 [ ~5.8] \\
17 & J195006+373236\tnote{$\ddagger$}  &  $<-$2.83       &  $<$ 180 &          7.0 &    22.7 [ 12.6] \\
18 & J195946+415143\tnote{$\ast$}      &  $<-$2.76       &  $<$1400 &          7.6 &    24.7 [ ~1.8] \\
19 & J195949+415324\tnote{$\ast$}      &  $<-$2.56       &  $<$ 810 &          7.6 &    24.7 [ ~3.0] \\
20 & J200008+415000\tnote{$\ast$}      &  $<-$2.91       &  $<$2040 &          7.6 &    24.7 [ ~1.2] \\
21 & J200011+415226\tnote{$\ast$}      &  $<-$3.05       &  $<$3820 &          7.6 &    24.7 [ ~0.6] \\
%%---------------------------------------%%
\hline
\label{table_uss}
\end{tabular}
\begin{tablenotes}
\item[] Notes. --- All the parameters are denoted the same way as in Table~\ref{table_gp}.
\item[$\dagger$,$\ddagger$,$\ast$] --- Targets marked with corresponding
symbols were observed together in common pointings.
\end{tablenotes}
\end{threeparttable}
\end{table*}
%%%=========================================================================

%%%=========================================================================
%%%%%%%
\subsection{Observations}
Observations were conducted at 327\,MHz using the Ooty radio telescope
\citep[ORT;][]{Swarup71}. The telescope has an offset, long parabolic
cylindrical reflector mounted equatorially, with a physical area of
15900\,m$^2$ and on-sky beam width of 1.75\mdeg\ and 6$^\prime$ in
east-west and north-south, respectively. 
While the steering in the east-west direction is mechanical, the beam
is steered electronically in the north-south direction. The effective
collecting area is estimated to be 55\% of the projected physical area
in the declination range of $-45$\mdeg\ to $+45$\mdeg. Outside this
range, the sensitivity drops rapidly with declination. The telescope
is receptive to only a single linear polarization (in north-south direction).
\par
Each of the sources in the above two samples was observed twice. The
first round of observations of the GP sample sources were
typically 5 or 10\,minutes long. The durations were estimated using
the expected flux densities at 327\,MHz such that the pulsed signals
from any associated pulsars would be detectable with a high significance
of 25$\sigma$. The second round of observations of this sample were
typically 30 or 45 minutes long. Observing durations of the
the USS sample sources in two rounds were 5 and 10 minutes, respectively.
In each observing session, raw voltage sequence was recorded at the
Nyquist rate (with 8-bit sampling) for a 16\,MHz wide band, centered at
326.5\,MHz using the new pulsar receiver PONDER \citep{Naidu15}.
%%%%%%%
\subsection{Search Processing}
The pulsar population is known to have a fairly steep radio spectrum
with an average $\alpha$ of $-1.4\pm1.0$ \citep{Bates13}, and the ones
with the steepest spectra ($\alpha<-2.5$) tend to be the fastest rotating
MSPs \citep{Kuniyoshi15,Kondratiev16,Frail16,Pleunis17,Bassa17}. To minimize
dispersion smearing across individual frequency channels and retain full
sensitivity for MSPs (and even sub-millisecond pulsars), we have employed a
semi-coherent dedispersion scheme implemented in \texttt{cdmt} \citep{BPH17}.
This search scheme is detailed in \citet{Bassa17} and uses the GPU accelerated
\texttt{cdmt} software to coherently dedisperse the raw voltage input data,
GPU accelerated incoherent dedispersion based on the \textsc{dedisp}
library \citep{Barsdell12} and tools from the PRESTO pulsar search
package \citep{RansomThesis}, including a GPU accelerated version of
the frequency-domain acceleration search technique \citep{Ransom02}.
\par
For each of the observations, we coherently dedispersed the raw voltage
sequence to 40 evenly spaced trial dispersion measures (DMs) in the
range 2.5$-$197.5\,\ppcc\, (both values inclusive) using \texttt{cdmt}
and recorded coherently
dedispersed data with 256 spectral channels (i.e., with channel widths
of 62.5\,kHz) and 16\,$\upmu$s sampling time
to disk. These channelized data were then incoherently
dedispersed around the corresponding coherent DM trials in steps of
0.01\,\ppcc, providing a uniform coverage of the DM range 0--200\,\ppcc\
in 20,000 incoherent trial DMs.
Similarly, data were coherently dedispersed to another 40 evenly
spaced trial DMs spanning the range from 210$-$990\,\ppcc\ with a
step size of 20\ppcc, and the
coherently dedispersed data were recorded with 256 channels and 64\,$\upmu$s
time resolution. The incoherent trial DMs were chosen in steps of 0.04\,\ppcc,
covering the range 200--1000\,\ppcc\ in another 20,000 steps.
The above configurations limit the dispersive smearing to a maximum of
about 40\,$\upmu$s and 150\,$\upmu$s in the lower and higher DM ranges, respectively,
retaining the sensitivity to even sub-millisecond pulsars subject to the
un-correctable smearing due to interstellar scattering.
\par
Each dedispersed time-series was searched for periodic signals using
\texttt{accelsearch} from the pulsar search and analysis software
\textsl{PRESTO} \citep{Ransom02}. For 5 and 10 minutes long observations,
we fixed the parameter \texttt{zmax} to 256, while for 30 and 45 minutes
long observations we used a value of 1024. These values imply that we have
searched for average accelerations of about 213 and 42\,m\,s$^{-2}$
of a 1000\,Hz signal, for observing durations of 10 and 45 minutes,
respectively. For each observation, 400 best pulsar candidates (200
each from the two DM ranges) were folded and the diagnostic plots were
examined by eye. The search pipeline was successfully validated using
several control pulsars.
\par
We also examined the faint candidates that were found to have consistent
periods (within 0.01\% of each other) and DMs in multiple observations
of same target fields. For such candidates, the data were folded and
the corresponding diagnostic plots for all observing sessions of a
particular target field were examined together.
\section{Results}
\begin{figure}
\begin{center}
\includegraphics[width=0.48\textwidth,angle=0]{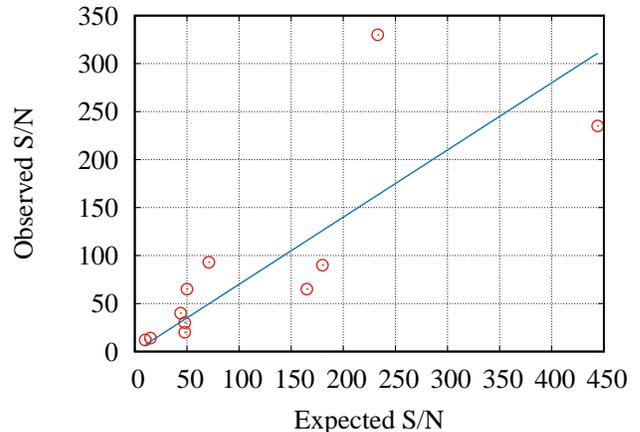}
\vspace*{-5mm}
 \caption{Expected versus observed signal-to-noise ratio (S/N)
for the known pulsars that were covered in the survey.}
 \label{fig_sbyn}
\end{center}
\end{figure}

Our survey did not yield any new pulsars. Several faint candidates
(6$-$8$\sigma$) that appeared to be potential pulsars turned out to be
false alarms in follow-up observations. In addition to the control
pulsars that were observed to validate the pipeline (PSRs~B1937+21, B1820$-$30A,
B1820$-$30B and B2002+31), we also detected several known pulsars which
happened to be in the beam or primary side-lobes (PSRs~J1908+0734, B2111+46,
B1844$-$04 and B1919+21) by chance or design. We have used the control pulsar
detections to make a realistic estimate of the achievable sensitivity
during our observations. The expected
signal-to-noise ratios (S/N) for the detections of above known pulsars
are plotted against the observed ones in Figure~\ref{fig_sbyn}. For
estimating the expected S/N, we have assumed an aperture efficiency of
55\% for the projected physical collecting area and a receiver temperature
of 150\,K. We also took into account the observed pulse-widths as well as
any offset in position from the beam center (a detection from a side-lobe
is not included). We also
estimated the direction dependent sky temperature using an all sky map
extrapolated to 325\,MHz \citep[for details, see][though here we use the
value at the beam center unlike the weighted average across the elongated
beam therein]{Maan17}. The observed
S/N is generally affected by inter-stellar scintillation, however,
we assume that observations of different pulsars as well as multiple
observations of individual pulsars provide us an average trend
between expected and observed S/N that is decided by the achievable
sensitivity. A straight line fit (see Figure~\ref{fig_sbyn}) suggests
that the achievable sensitivity is about 70\% of what is suggested
by the radiometer equation. It is worth emphasizing here that the
above sensitivity degradation factor (70\% or 0.7) includes the aspects
such as the effective bandwidth, which is typically smaller due to the
effect of filter roll-off on the band edges, and reduction in sensitivity
due to low-level radio frequency interference.
\par
The 10$\sigma$ upper limits on flux densities of a periodic signal with 10\%
duty-cycles from the target fields are listed in Tables~\ref{table_gp} and
\ref{table_uss} (see the parameter $S_{\rm min}^{\rm pulsed}$). These upper limits
have taken in to account the above deduced sensitivity degradation factor.
While scintillation could have prohibited detection of a few
of these sources, our tight upper limits suggest that these
sources, as a population, are not observable radio pulsars at 327\,MHz.
%
%%%%-------------------------------------------------------------------------
%%%-------------------------------------------------------------------------
%%%-------------------------------------------------------------------------
%%
%
\section{Discussion}
The natural question follows: if not pulsars, what could these sources be?
Many of the earlier pulsation surveys of image-based targets, e.g.,
\citet{Damico85,Kaplan00,Crawford00}, concluded that most of their target
sources were possibly extragalactic.
Only a few categories of extragalactic sources (e.g., extended emission from
radio halos and relics in merging galaxy clusters, HzRGs) are known to exhibit
spectral indices as steep as $-$1.5. Within our galaxy, only pulsars are known
to have such steep radio spectra.
As noted earlier, \citet{DeBreuck00} and \citet{deGasperin18} have
reported an excess of steep-spectrum, compact sources in the Galactic plane.
For Galactic latitudes $\abs{\emph{b}}$$\leq$10\mdeg, \citet{deGasperin18} quantify
the excess at 28\%. For the Galactic latitude range of our GP sample, i.e.,
$\abs{\emph{b}}$$\leq$2\mdeg, the excess is even higher and appear to be as much
as 50\% \citep[][see their Figure 15]{deGasperin18}, and these are clearly of
Galactic origin. If we assume that half of the remaining 50\% steep-spectrum
sources in this Galactic latitude range are extragalactic, 75\% of the targets
in our GP samples can be expected to be Galactic. Unless we are looking at a
previously unknown class of steep-spectrum sources in our Galaxy, majority of sources in
our GP sample should still be pulsars but somehow missed in our searches.
A few, specific scenarios explaining such non-detections of presumably associated
pulsars have been
considered by many authors, including \citet{Crawford00} and \citet{deGasperin18}.
Below we discuss if, particularly in the context of our survey, these scenarios
could possibly lead to non-detection of the underlying pulsar population in these
steep spectrum sources.
\subsection{Pulsars in very tight binary systems}
Signals from pulsars in tight and relativistic binary systems experience
varying doppler shifts as a function of orbital phase which could make
their detection difficult. However, we have searched for accelerations
up to more than 200\,m\,s$^{-2}$ in our shorter duration observations.
%\par
That should be sufficient to detect all currently known relativistic double neutron-star binaries. 
Even the accelerations of the top 4 such binaries are within our search range.
PSR~J1906+0746 \citep{2015ApJ...798..118V} achieves a maximum acceleration of $\sim$95\,m\,s$^{-2}$;
the MSP in the double pulsar, PSR~J0737$-$3039A, was detected at 99\,m\,s$^{-2}$ and reaches up to 250\,m\,s$^{-2}$ \citep{2007mru..confE..92E};
the 1.88-hr orbit of PSR~J1946+2052 \citep{2018ApJ...854L..22S} imparts a maximum acceleration over 300\,m\,s$^{-2}$.
These three systems are all relatively circular. The highly eccentric orbit of PSR~J1757$-$1854 \citep{2018MNRAS.475L..57C} means that while it was detected at 32\,m\,s$^{-2}$, its maximum encountered acceleration is in excess of 600\,m\,s$^{-2}$.
\par
The acceleration search methods (like the one employed by \texttt{accelsearch})
are most efficient when the duration of the signal is less than or around
one-tenth of the orbital period \citep{Ransom01}. It might imply that our search was not
very sensitive to orbital periods shorter than an hour or so. However,
we note that our upper limits are nearly an order of magnitude better
than the expected flux densities, and we would have easily detected
the signals in much smaller temporal sections at appropriate acceleration
values. Hence, such a scenario is unlikely to have hindered detections
of any underlying pulsars.
%%
%%%%%%%
%%%-------------------------------------------------------------------------
\begin{figure}
\begin{center}
\includegraphics[width=0.38\textwidth,angle=-90]{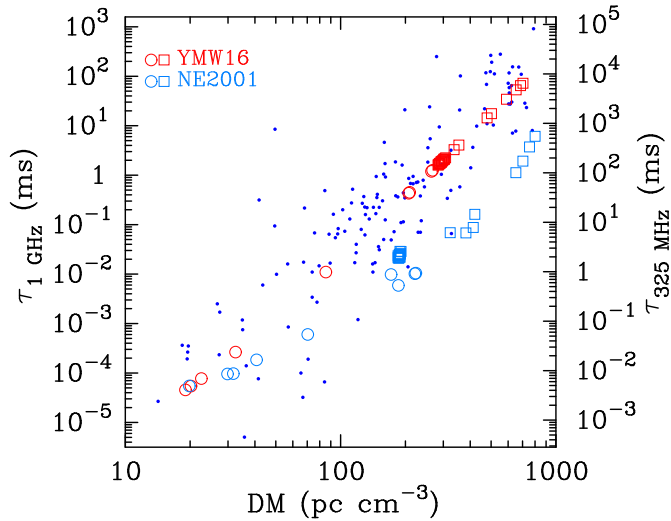}
\caption{\emph{The worst-case scatter-broadening:} The blue dots are the
scatter-broadening measurements for 148 known pulsars.
The open squares and circles indicate the \emph{maximum}
DMs (i.e., assuming the sources to be at the outer edges of the Galaxy)
and corresponding characteristic scatter-broadening at 1\,GHz
for the targets in our GP and USS samples, respectively. The corresponding
scatter-broadening at 325\,MHz is indicated by the right hand side
vertical axis assuming a $\tau\propto\nu^{-4}$ relationship.
The red and
blue colored symbols indicate that the quantities are estimated using
the YMW16 and NE2001 models, respectively.}
 \label{fig_scat}
\end{center}
\end{figure}
%%%-------------------------------------------------------------------------
\subsection{Highly scattered pulsars}
Scattering in the ionized interstellar medium results in broadening of
pulsed periodic signals from pulsars. Scattering-induced broadening generally
has steep dependence on the observing frequency and DM of the source, with
the effect becoming highly pronounced at lower frequencies and higher DMs.
Depending on the observing frequency and DM, the scatter-broadened
pulse-width could become comparable, or even larger than the pulse period,
in which case detection of the periodic signals becomes very
difficult or even impossible.
\par
To investigate if scattering could have affected detection of some of the
potential pulsars in our survey, we have estimated the maximum Galactic
DM towards each of the target fields using the electron density models by
\citet[][hereafter YMW16]{YMW16} and \citet[][hereafter NE2001]{CL02}.
These maximum DMs and the corresponding estimates of the characteristic
scatter-broadening ($\tau$) at 1\,GHz and 325\,MHz are shown in Figure~\ref{fig_scat},
along with the available measurements from 148 known pulsars
\citep[][measurements at different frequencies are scaled to 1\,GHz]{Lohmer01,Lohmer04,Lewandowski15,KK17,Geyer17,KK18a,KK18b}.
NE2001 seems to underestimate the scatter broadening for majority of the
targets while YMW16 estimates use the empirical relationship from \citet{KK15}
and naturally follow the average trend when
compared with the available measurements. Considering the YMW16 estimates,
for majority of the targets in our GP sample the \emph{maximum} Galactic contribution
to DMs is around 300\,\ppcc, and for the others its in the range 500--700\,\ppcc.
For the USS sample, the \emph{maximum} DMs for majority of the targets are below
100\,\ppcc, and those for the rest are between 200--300\,\ppcc.
The right vertical axis of Figure~\ref{fig_scat} indicates the
scatter-broadening at 325\,MHz assuming a $\tau\propto\nu^{-4}$ relationship.
We note that the distances corresponding to the maximum Galactic DMs are about
20$-$25\,kpc for each of the sources in our GP sample. So, if all the targets are
really at the \emph{extreme outer edges} of the Galaxy in their respective
directions, then indeed our survey would not be sensitive to many of
the GP-sample sources for rotations periods shorter than about 50$-$100\,ms.
However, the DM estimates from YMW16 model could easily have uncertainties
of the order of 50\% or even more. If we assume the DMs to be half of those suggested
by YMW16, then 19 of the 25 sources in our GP sample would have scatter-broadening
less than or around 10\,ms.
Furthermore, if these sources are only a few kpc away, say 3\,kpc, then the
scatter-broadening would be much less than or around 10\,ms for all of the sources
in the GP sample. Moreover, at this distance, considering the 50\% uncertainties
on the DMs, the scatter-broadening could be potentially less than a millisecond
for majority of the GP sample targets. On the other hand, if we consider the DMs
to be underestimated by a factor of 2, then more than 50\% of the GP sample sources
would not be detectable even from a distance of 3\,kpc if their spinning periods
are shorter than 100\,ms.
\par
Most of the targets in USS sample are still expected to have scatter-broadening
much less than 1\,ms, even if they happen to be at outer edges of the Galaxy,
implying no loss of sensitivity due to scattering
for these sources even for sub-ms rotation periods.
\par
Measurements from known pulsars show more than two orders of magnitude scatter of pulse-broadening around a power-law trend (see Figure~\ref{fig_scat}), and even anomalously higher scattering for some. If all of these steep spectrum sources happen to exhibit anomalously high scatter broadening, then signals with short periods would not be detectable even at 1.4\,GHz.
\subsection{Aligned or nearly aligned rotators}
The misaligned rotation and magnetic axes, and the beaming of radio emission around
the latter, gives rise to the observed pulsed periodic signals from pulsars.
However, there is some evidence that the two axes approach towards
alignment with age, on typical timescales of $10^7$~years \citep{TM98}.
In a scenario where the magnetic inclination angle approaches the
half opening angle of the radio beam, but
still larger, the primary observable effect
would be the increased pulse duty cycle. Detection of periodic signals
with large duty cycles requires more sensitive observations
than that of narrow duty-cycle signals, since the overall flux is smeared out
over more bins, and harmonic summing in the Fourier domain search becomes less effective.
In fact, less than 1\% of known pulsars have pulse widths wider  than half the
rotation period. For rotation periods longer than 100\,ms, only
a few pulsars are known for which the \emph{intrinsic} radio
emission spans a significant fraction of the period
\citep[e.g., PSR~B0826$-$34, PSR~J1732$-$3131;][]{AL81,MAD12}.
However, our observations were of such sensitivity that 
even if the pulsed emission spanned 90\% of the period,
signals from a majority the periodicities would have been detected.
So, our target sources, as a population, are unlikely to be just
large duty-cycle pulsars. 
\par
\begin{figure}
\begin{center}
%\fbox{
\includegraphics[clip,trim=0 100mm 0 5mm width=0.8\columnwidth]{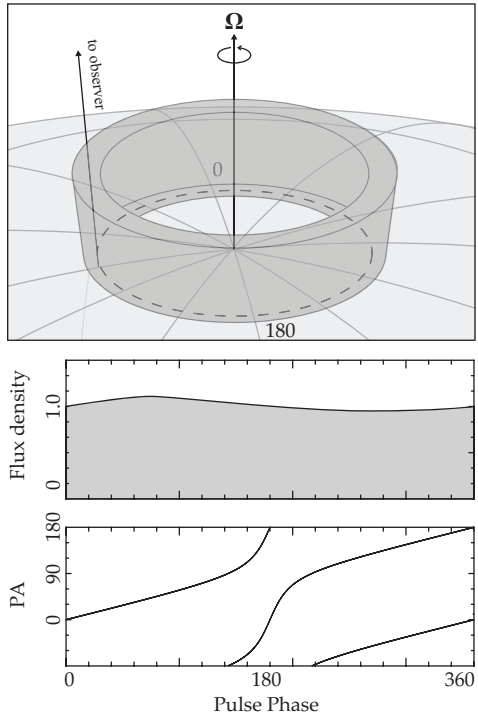}
 \caption{Illustration of an aligned rotator with $\alpha=5\mdeg$ (top sub panel), where the expected fraction of the flux that is pulsed is low (middle) but the linear polarization changes throughout the period (bottom).}
 \label{fig_pa}
\end{center}
\end{figure}

Once the magnetic inclination angle becomes comparable or smaller
than the half opening angle of the radio beam, a good fraction
of the radio emission would appear as \emph{continuous} emission 
(cf. top and middle panels of Figure~\ref{fig_pa}).
As the typical pulse shape from most pulsars is far
from a ``top-hat'', these intrinsic variations would cause some modulation in the observed
intensity over the pulsar period, even for a nearly aligned rotator.
However, only a fraction of the total flux density would be apparent
as \emph{pulsed}, and that too with large duty cycles.
One known example of such a system is PSR~J0218+4232, where only about half of the total
radio flux is pulsed \citep{Navarro95}.
The increased baseline does not, by its own right, make pulsar searching harder. 
A successful detection in such a search
is determined only by the signal-to-noise of the pulsed signal. 
And given our sensitive observations, in a number of cases we could have measured
such small variations.
Assuming a duty cycle of 70\%, in the last columns of Tables~\ref{table_gp}
and \ref{table_uss}, we place upper limits on pulsed fractions of the
target-source flux density. We can generally rule out pulsed
flux densities higher than a few tens of percents.
Probing a scenario where the pulsed flux is only a few per cent of
the total observed emission would require significantly deeper observations.
\par
A perfect alignment suggests a cessation of any ``pulsar'' action as an observable phenomenon.
However, for an asymmetric emission beam that pulsars are generally
known to have \citep[cf.~the beam mapped for PSR~J1906+0746;][]{2013IAUS..291..199D},
the observable implications are identical to those described
for nearly aligned rotators above.
%%%-------------------------------------------------------------------------
%%%-------------------------------------------------------------------------
%%
\par
While the total intensity in an aligned rotator may be steady, other properties of the emission \emph{do} vary with pulse phase. Throughout the pulse period, the line of sight will steadily traverse 360\mdeg\ in azimuth, around the magnetic pole. In the rotating vector model \citep{RC69}, this translates to a full rotation of the angle of linear polarization. Even under invariant total intensity, position-angle sweep may be visible in sources with some fractional linear polarization (Fig.~\ref{fig_pa}). The same interstellar medium that inflicts the frequency dispersion to the pulsar signal will also Faraday rotate. The magnitude of both effects is unknown but could be dealt with similarly; through a search in both DM (this work) and rotation measure (RM). 
Using the RM-DM relation \citep{1989MNRAS.237..533L} would help limit the range of trial RMs per trial DM. From the polarization-angle swings found after  applying a fast folding algorithm, aligned rotators could be disentangled from background sources.
Alternatively, a time series formed from the values of the polarization angle in each sample could be searched in a Fourier transform. With the period, DM and RM known, deep follow up at higher frequencies, where the narrowing beam may no longer cover the entire pulse period, could provide insights into the pulsar geometry.
\par
Since ORT is receptive to only a single linear polarization, the differential Faraday rotation of a linearly polarized signal manifests itself in the form of spectral intensity modulation within the observation bandwidth. This spectral modulation can be exploited to deduce RM and linear polarization properties \citep{RD99,Maan15}. However, this is possible only for reasonably high RMs and significantly linearly polarized sources, and the above mentioned search involving RM would require computing time several orders of magnitude longer than that spent in the current search limited to DM, period and acceleration domains.
\par
In very specific cases, the observation bandwidth might sample only a small and near-bottom fraction of the above Faraday rotation induced spectral intensity modulation, which could potentially lead to a non-detection. However, such a non-detection would essentially need the signal to exhibit nearly 100 percent linear polarization, appropriately low RM, and a near constant polarization position angle (PPA). For most of the sources in our GP sample, the upper limits on linearly polarized components are 20$-$30\% of the total flux density (NVSS). The small linear polarization fraction may be intrinsic to the sources or might be indicating depolarization due to well known large PPA swings exhibited by pulsars or scattering. In either of the cases, ORT's reception to only a single linear polarization is unlikely to affect the potential detection.
\section{Summary}
Summarizing, we have presented a search for pulsars in 44 steep spectrum sources, employing a semi-coherent dedispersion scheme. While our survey did not yield any new pulsars, we argue that majority of the sources in our samples should actually be pulsars or a new category of \emph{Galactic} sources. Non-detection of any pulsars from our GP sample of 25 sources suggests that even at 327\,MHz the excess of steep spectrum sources in the Galactic plane can not be accounted for by a conventionally observable pulsar population.
Interstellar scattering could have affected our searches only if majority of the
sources in our sample happen to be pulsars with ms or a few 10s of ms rotation
periods and they are either located at \emph{extreme outer edges} of the Galaxy or
exhibit anomalous scattering.
A fraction of the MSPs located a few kpc away could also have been missed.
We have also discussed in detail the scenario that these sources are in fact aligned or nearly aligned rotators and propose methods to probe such a situation further. \emph{Deep} searches at higher frequencies (a few GHz) will in general be helpful in uncovering interesting pulsars, in case these steep spectrum sources harbor anomalously scattered or (nearly) aligned rotators.
\acknowledgements
YM is grateful to Prabhakar Tiwari for providing access to the data presented in \citet{Tiwari16}. We acknowledge the kind help and support provided by the members of the Radio Astronomy Centre, Ooty, during these observations. ORT is operated and maintained at the Radio Astronomy Centre by the National Centre for Radio Astrophysics.
YM and JvL acknowledge use of the funding from the European Research Council under the European Union's Seventh Framework Programme (FP/2007-2013)/ERC Grant Agreement no. 617199. This work has made use of the DRAGNET GPU cluster, as well as the Dutch national e-infrastructure with the support of SURF Cooperative. Computing time was provided by NWO Physical Sciences (project no. 16268). CB acknowledges support from the European Research Council under the European Union's Seventh Framework Programme (FP7/2007$-$2013)/ ERC grant agreement No. 337062 (DRAGNET; PI: Hessels). MAK and BCJ acknowledge support from Department of Science and Technology Extra-mural grant DST-SERB EMR/2015/000515.
This research has made use of the NASA/IPAC Extragalactic Database (NED) which is operated by the Jet Propulsion Laboratory, California Institute of Technology, under contract with the National Aeronautics and Space Administration
%%
%\software{PRESTO \citep{Ransom02}, cdmt \citep{BPH17}, \textsc{dedisp} library \citep{Barsdell12}}
%%
%%%%%%%%%%%%%%%%%%%%%%%%%%%%%%%%%%%%%%%%%%%%%%%%%%
%%%%%%%%%%%%%%%%%%%% REFERENCES %%%%%%%%%%%%%%%%%%
%%%%%%%%%%%%%%%%%%%%%%%%%%%%%%%%%%%%%%%%%%%%%%%%%%

%%%%%%%%%%%%%%%%%%%%%%%%%%%%%%%%%%%%%%%%%%%%%%%%%%
%%%%%%%%%%%%%%%%%%%%%%%%%%%%%%%%%%%%%%%%%%%%%%%%%%
%%%%%%%%%%%%%%%%%%%%%%%%%%%%%%%%%%%%%%%%%%%%%%%%%%
%%
%%
\end{document}